# Embedded Sensor System for Early Pathology Detection in Building Construction

Santiago J. Barro Torres, Carlos J. Escudero Cascón

**1 Department of Electronics and Systems, University of A Coruña**
**A Coruña, 15071 Campus Elviña, Spain**
***sbarro@udc.es***

**2 Department of Electronics and Systems, University of A Coruña**
**A Coruña, 15071 Campus Elviña, Spain**
***escudero@udc.es***

## Abstract

Structure pathology detection is an important security task in building construction, which is performed by an operator by looking manually for damages on the materials. This activity could be dangerous if the structure is hidden or difficult to reach. On the other hand, embedded devices and wireless sensor networks (WSN) are becoming popular and cheap, enabling the design of an alternative pathology detection system to monitor structures based on these technologies. This article introduces a ZigBee WSN system, intending to be autonomous, easy to use and with low power consumption. Its functional parts are fully discussed with diagrams, as well as the protocol used to collect samples from sensor nodes. Finally, several tests focused on range and power consumption of our prototype are shown, analysing whether the results obtained were as expected or not.

## 1. Introduction

Over the last years a growing interest in security in the field of building construction has been noticed. The knowledge of distortions and movements caused by the structures at the right time makes it possible to assess their tension and, consequently, improve workers' security.

The techniques traditionally used in the inspection of structures are very basic, mostly centered on having an operator watching out for damages present in the materials (fissures in the concrete, metal corrosion...). Sometimes, the structure is hidden or is difficult to reach. Additionally, access to the structure can be dangerous, as is the case with bridges. All these problems complicate the examination process. Therefore, it is necessary to have alternative means of detecting pathologies [1, 2, 3].

Advances in microelectronics make it possible to design new systems for carrying out technical inspection of works. Nowadays, it is possible to obtain embedded systems with a high degree of integration, processing, storage and a low consumption at an affordable price. On the other hand, sensor networks have evolved up to the point that it is possible to have a series of sensors sharing information and cooperating to reach a common goal. Thus, this new generation of intelligent sensors is beginning to look like a suitable technology for pathology detection. The saving in maintenance costs in the near future would make it possible to recover the value of the initial investment, by avoiding hiring technical inspection services.

The boom of wireless technologies has also reached the sphere of sensor networks, with technologies such as ZigBee [4, 5, 6], which lets us interconnect a group of sensors in an ad-hoc network without a default physical infrastructure or a centralized administration [7]. For that reason, this technology is very suitable for this application.

In this article the design and implementation of a pathology detection network based on embedded systems, sensors and the ZigBee technology is presented, satisfying the following requirements, as for example:

- *Ease of Use*. The network must be able to configure itself, without human intervention, reducing maintenance costs.
- *Fault tolerance*. In case one of the intermediate nodes fail, the network would look for alternative roots, so as not to leave any node isolated.
- *Scalability*. The network should be as extensible as possible, to place the sensor nodes located in areas potentially far away from the building.
- *Low consumption*. Sometimes it will be difficult or impossible to feed the nodes directly from the electric

**IJCSI**

network, so it is necessary to equip them with a battery they will have to take full advantage of.

- *Flexibility*. The frequency with which samples are taken will change with time, which means that periodicity must be an independently configurable parameter in every single node. In addition, modification of the periodicity must be possible at any moment, even when the node is sleeping.

The system presented in this article is made of a series of sensor nodes specialized in the detection of pathologies, distributed along the structure of a building. These nodes communicate wirelessly through a ZigBee mesh network, which makes expanding the network easier, as there is no default structure. Moreover, management becomes easier, too. One of these nodes, formally called coordinator, is in charge of gathering and storing the samples sent by sensor nodes at a configurable periodicity, for their use in subsequent studies. Sensor nodes, in turn, are fed by a battery, and can thus operate autonomously, but at the cost of using a power saving design.

The article structure is as follows: The second section describes the problem and the technical solution adopted. The third section presents the logical design of the system, showing its different parts and explaining how they operate. The fourth section deals with implementation, including commented photos of the prototype built. The fifth section shows the test results performed. Finally, the sixth section is dedicated to the conclusions.

## 2. Problem Statement and Technology

The choice of sensors depends on the physical phenomenon to be measured. In this case, the most suitable sensors to perform pathology detection are strain gauges [8], potentiometric displacement sensors and temperature catheters [9, 10].

Strain gauges are used to measure the deformation level of materials such as concrete and steel. Its operation is based on the piezoresistive effect, which means that its internal resistance changes when is deformed by an external force. As the voltage variations obtained are very small (less than 1 mV), it is necessary to add extra circuitry to condition the signal prior to reading its value: Amplification, noise filtering, etc. [11] Heating the gauge several minutes before sampling is another important restriction. On the other hand, potentiometric displacement sensors are used to measure the movement suffered by the structure with respect to its anchors. An important difference between the two sensors mentioned is that whilst strain gauges are not reusable, displacement sensors are, because gauges are attached or embedded within the structure. Another kind of sensor to take into account is temperature catheters, which allows concrete monitoring in its first stages after the building construction has been completed.

ZigBee is a specification for a suite of high level communication protocols using small, low-power digital radios based on the IEEE 802.15.4 standard for wireless personal area networks (WPANs). In turn, IEEE 802.15.4 specifies the physical layer and media access control for low-rate wireless personal area networks (LR-WPANs) [12].

The purpose of ZigBee is to provide advanced mesh networking functions not covered by IEEE 802.15.4. ZigBee operates in the industrial, scientific and medical (ISM) radio bands; 868 MHz in Europe, 915 MHz in the USA and Australia, and 2.4 GHz in most jurisdictions worldwide [4]. This technology is intended to be simpler, cheaper and to have lower power consumption than other WPANs such as Bluetooth.

In a ZigBee network there are three different types of nodes:

- *ZigBee Coordinator*: The coordinator forms the root of the network tree and might bridge to other networks.
- *ZigBee End Device:* Contains just enough functionality to talk to the parent node (either the coordinator or a router), and it cannot relay data from other devices. This relationship allows the node to be asleep a significant amount of the time thereby giving long battery life.
- *ZigBee Router:* Acts as an intermediate router, passing on data from other devices. Moreover, it might support the same functions as a *ZigBee End Device*.

Nowadays, ZigBee-based devices are easy to find, as many semiconductor and integrated circuit manufacturers have opted for this new technology:

- *Digi International*, leader in *Connectware* solutions, offers development kits for its module *XBee® & XBee-PRO® ZB RF* [13].
- *Rabbit*, an 8-bit microcontroller specialized company, has developed the arquitecture *iDigi™ BL4S100*, which consists of a *XBee-PRO® ZB* module with a *Rabbit® 4000* microcontroller, capable of acting as an intelligent controller or *ZigBee-Ethernet* gateway [14].
- *Ember*, a monitoring and wireless sensor network provider offers the *InSight Development Kit* [15],

which includes everything needed to create embedded applications in their radios *EM250/EM260*.

- *Crossbow*, one of the leading wireless sensor manufacturers, features several development kits that provide complete solutions in the development of such networks, including the *Professional Kit* [16].
- *Sun Microsystems* offers their Java-based *Sun SPOT* [17], formed by a processor, battery and a set of sensors.

# 3. Design

This section shows the architecture and models describing the presented system.

## 3.1 System Architecture

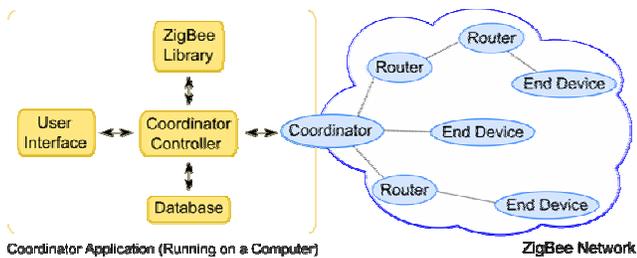

Fig. 1 System Architecture

The system comprises a set of nodes interconnected in a ZigBee network, as shown in Figure 1. This structure provides us mobility to place end devices anywhere in the building, as well as connectivity to collect samples:

- The Coordinator node works as gateway between the sensor network and the main station computer, where the Coordinator Application Software will be running, which provides the user an interface to manage the system
- Router Nodes extend the network coverage through the whole building. They need to be placed where can be powered without interruption.
- End Devices sample sensor at regular intervals, and then send the values obtained to the coordinator. When not in use, the end device enters a low power consumption mode, called 'sleep mode', helping to increase battery life.

The Coordinator and End Devices collaborate with each other, following the protocol explained below. This protocol is important to ensure that samples are collected properly.

## 3.2 Communications Model

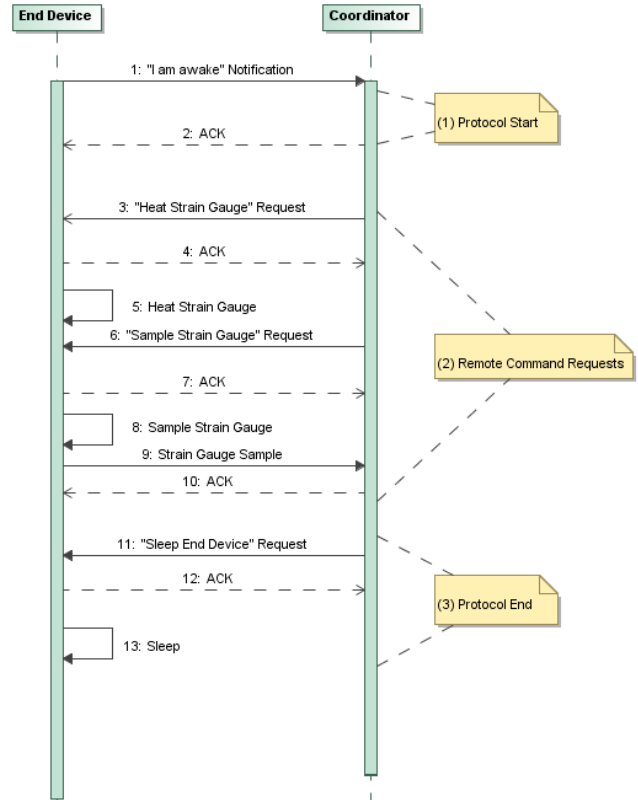

Fig. 2 Sample Colleting Protocol Sequence Chart

The Coordinator and End Devices interact using a special state message-based protocol (see Figure 2). The protocol starts when any of the End Devices of the network decides to awake, notifying the Coordinator such situation with the message *"I am awake"* (1). This makes the sample periodicity management easier, since each End Devices know when it is time to take the next sample. Once the protocol has been started, the Coordinator is completely free to send the End Device any remote commands requests (2). For example, heating the gauge, as this is one of the requirements of this type of sensors. When the gauge is ready, the Coordinator could ask the End Device for as many samples as necessary. This approach gives us great flexibility to adapt the protocol to the application needs, with minimal system design changes. Finally, the Coordinator asks the End Device to sleep immediately (3). This event marks the end of the protocol, which will be executed again when the End Device decides to send the next sample, according to the sampling periodicity settings.

The protocol requires processing capacity both in the Coordinator and the End Device, as illustrated in the next subsection.

**Coordinator Flow Chart (Figure 3):** The Coordinator waits for the message "I am awake", which is sent by one of the End Devices who owns the periodicity control. When this message arrives, the Coordinator asks the End Device to heat its strain gauge. After a while, the Coordinator asks the End Device to send one sample, which is stored in the Coordinator's database whenever it is received. Finally, the Coordinator puts the End Device to sleep.

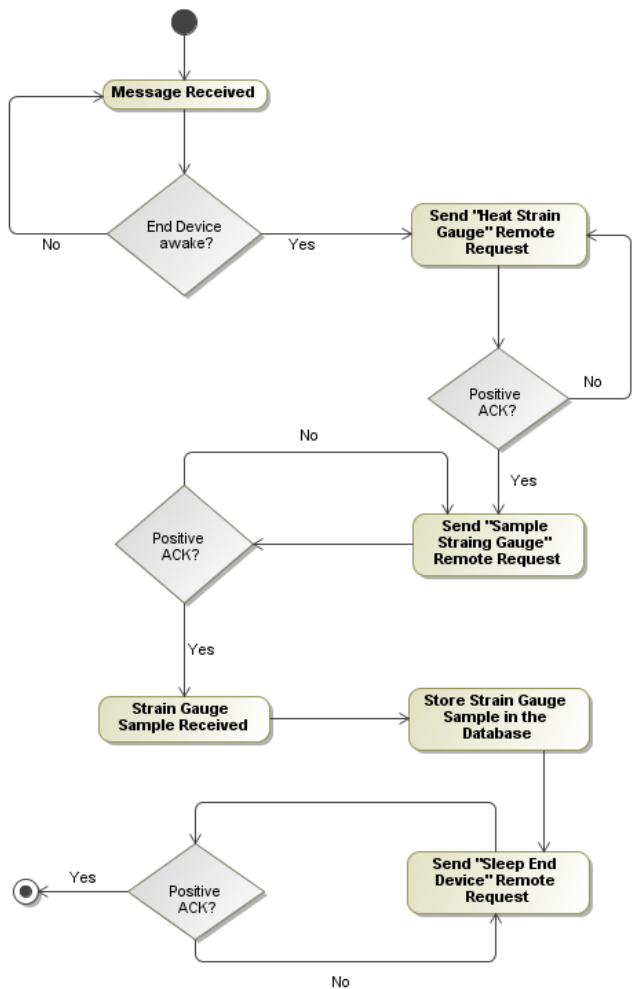

Fig. 3 Coordinator Flow Chart

**End Device Flow Chart (Figure 4):** The End Device has its own internal timer to know when to wake up, according to the sampling frequency settings. When it is time to wake up, the End Device notifies the Coordinator with the message "I am awake", and enters an idle state, waiting for remote request commands coming from the Coordinator. According to the example being shown, there are three possible commands: *Heat Gauge*, *Sample Gauge* and

*Sleep*. Whenever a remote command is received, the message type is checked and the appropriate action is executed. The protocol ends when the Coordinator sends a Sleep End Device Request.

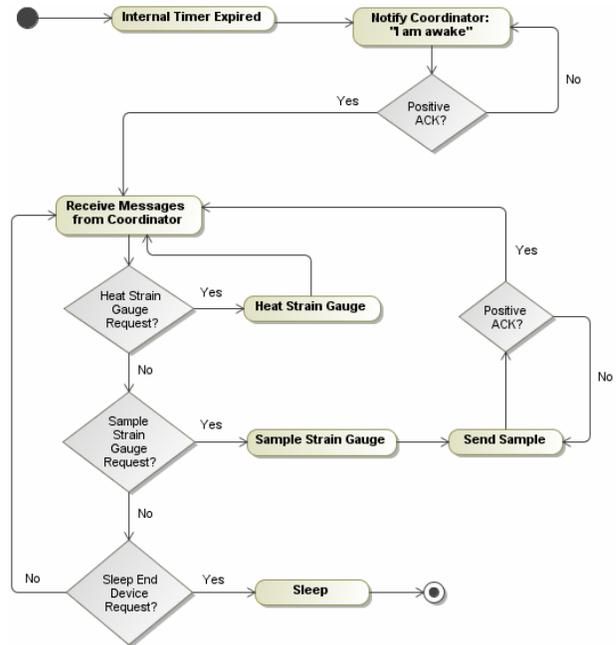

Fig. 4 End Device Flow Chart

### 3.3 Network Model

The network model is composed by three different submodels, in accordance with their responsibilities, as it was mentioned earlier:

- Coordinator Model
- Router Model
- End Device Model

**Coordinator Model:** The coordinator is responsible for collecting samples from the end devices, besides network management. There is only one in the entire network, and consists of the following elements:

- *User Interface.* Some of the operations performed on the system require user interaction, hence the need for an entry data interface.
- *XBee-API Communications Library.* It is an object model library specially designed to talk to XBee ZB modules in API mode.
- *Database.* It is used to store samples persistently, so that they can be analyzed or queried later.
- *Coordinator Controller.* Here lies the Coordinator core, where the protocol operations are performed.

IJCSI

**Router Model:** Routers are useful to extend the network coverage, enabling communication between the Coordinator and End Devices. This communication would not be possible or would be very difficult to establish because of the distance, presence of obstacles (walls, floors, ceilings, structures, etc.).

Although its presence on the network is optional, routers are usually distributed among several strategic points to extend coverage effectively.

**End Device Model:** Figure 5 shows the functional elements compounding an End Device:

- *Control Unit.* Here lies the End Device core, where the protocol operations are performed. Besides, it manages all other components and performs actions such as: *Sampling sensors, sending and receiving ZigBee messages, setting the alarm clock, etc.*
- *Alarm Clock.* Once configured, it is able to wake up the whole End Device circuitry. Sampling frequency is set here. It must be always powered.
- *ZigBee End Device Module.* Enables remote communication with the Coordinator through ZigBee technology. It must be always powered, as it may receive data at any time, even when the End Device is asleep.
- *Conditioning Circuitry.* It is responsible for adapting the signal obtained in the strain gauge (amplifying, filtering, ADC-converting, etc.) so the control unit can read its value.

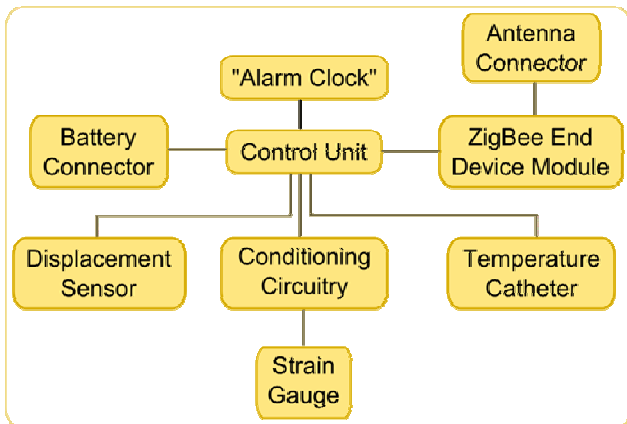

Fig. 5 End Device Functional Model

# 4. Implementation

## 4.1 Prototype Description

The hardware platform selected was *SquidBee* [18], an open mote architecture formed by *Arduino Duemilanove Board* [19], *Digi International XBee® ZB* [20] and a set of basic sensors (temperature [21], light and humidity [22] sensors), distributed by *Libelium* [23]. The main advantage of this platform is its great flexibility, as it allows us to build any node (Coordinator, Router or End Device) with very few hardware and firmware changes to its basic architecture [24].

The End Device's Unit Control has been implemented in the Arduino's Atmel ATmega168 microcontroller [25]. On the other hand, the XBee® ZB integrates the "*alarm clock*", since it is able to wake up at regular customizable intervals (formally, cyclic sleep [20]). Therefore, remote configuration is much easier, as XBee provides simple mechanisms to change remote variables from the Coordinator. Finally, the Coordinator Application is a Java-based Desktop Software running on a computer with the Coordinator connected to one of its USB ports. *XBee-API Library* [26] has been used to implement such application.

## 4.2 Coordinator Implementation

Here is the component list for the Coordinator, as shown in Figure 6:

- *Rigid case.* Protects the components and the circuitry.
- *USB cable.* The connection between the Coordinator and the Computer is established by a USB connector. The USB also powers this Node.
- *XBee® ZB RF Module.* Provides the End Device with ZigBee connection. Must be set up with the Coordinator *firmware*.
- *2.4 GHz Antenna.* An external antenna to connect to XBee, using a U. FL proprietary connector from *Hirose Electronic Group* [27].
- *Arduino Board (without microcontroller).* Since the Coordinator Application is running on the computer, the microcontroller is not needed anymore.
- *XBee Shield.* Enables ZigBee connection to the Arduino Board. USB connection must be selected [28].

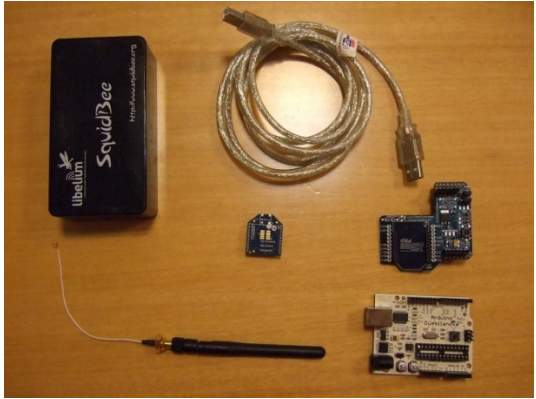

Fig. 6 Coordinator components

### 4.3 Router Hardware Implementation

In this case, the component list remains almost the same. Therefore, only different and new components are highlighted:

- *USB charger.* Unlike the Coordinator, Routers are not connected to a Computer, so a USB charger is needed to plug the Node to the power supply.
- *XBee® ZB RF Module.* Despite being physically equal, a different *firmware* is needed. Router *firmware* must be set up in the module [29].

The final assembled node is shown in Figure 7.

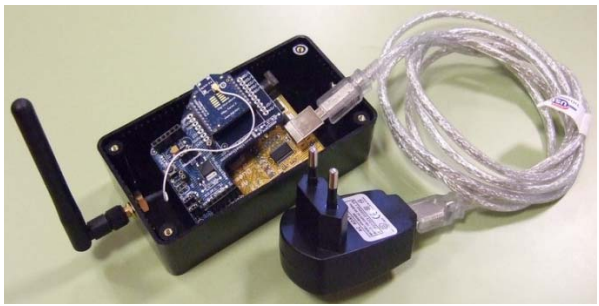

Fig. 7 Assembled Router

### 4.4 End Device Hardware Implementation

Again, only different and new components are highlighted. The components are shown in Figure 8:

- *Rechargeable Lithium Battery.* Since the End Device needs full autonomy, it must be powered by a Battery (of 1100 mAh in our case)
- *Arduino Board with Atmel ATmega168.* The Arduino Board must have its microcontroller with our protocol implementation loaded in the

memory.
- *XBee Shield* configuration is slightly different from the others. First, an extra connection is needed to awake the microcontroller from XBee [29]. Second, the XBee connection must be selected (mind the USB connection was the one selected previously) [28].
- *XBee® ZB RF Module.* End Device *firmware* must be set up [24].
- *Data Acquisition Board.* It is the special circuitry shown in Figure 9, which is equipped with several signal conditioning components (Analog to Digital Converters or ADCs, among others) used to connect strain gauges, potentiometric displacement sensors, temperature catheters and so on. Communication between the End Device and the Data Acquisition Board is performed through a Serial Port Connection, using a specific set of commands.

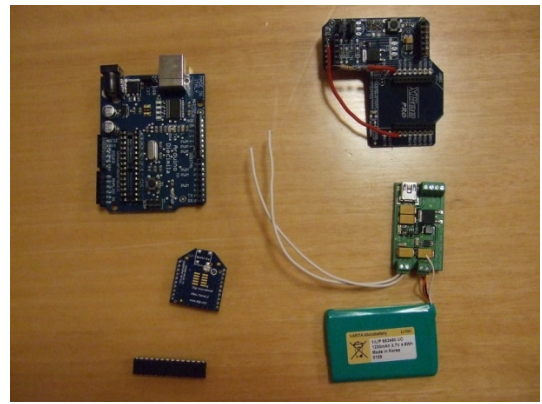

Fig. 8 End Device components

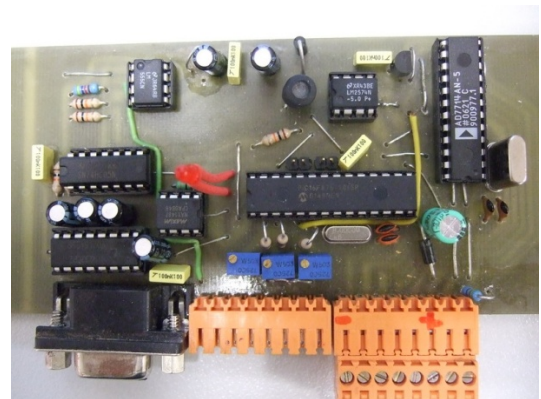

Fig. 9 Data Acquisition Board

As we will see in the Testing Scenario, we take temperature samples from the Data Acquisition Board, which has one temperature sensor [21] attached to one of its multiple input channels. The communication between the End Device and the Data Acquisition Board is performed through a RS-232 or Serial Connection. Since there are no more Serial Connectors available in the Arduino Board (apart from the one establishing the communication with the ZigBee Network), two virtual Serial Port Connectors were created, using a special circuitry (shown in Figure 10) and an Arduino Library called *SoftwareSerial* [30]. One of those ports establishes the communication, and the other one enables the log output, which is helpful when performing debugging tasks.

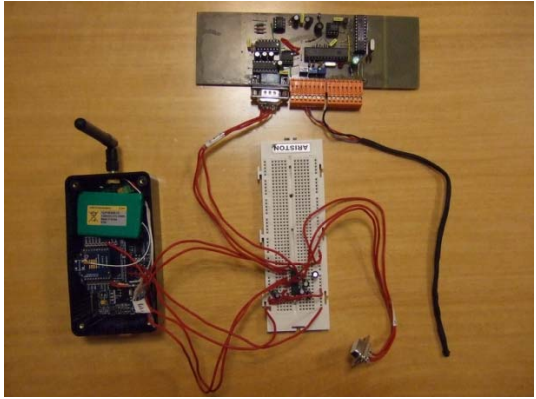

Fig. 10 Fully assembled End Device

This special circuitry converts Arduino TTL voltage levels into RS-232 values, since a RS-232 connector is needed to talk to both the Data Acquisition Board and Log Software. The electronic scheme, shown in Figure 11, is based on the ADM232L chip [31], which supports up to 2 Serial Ports.

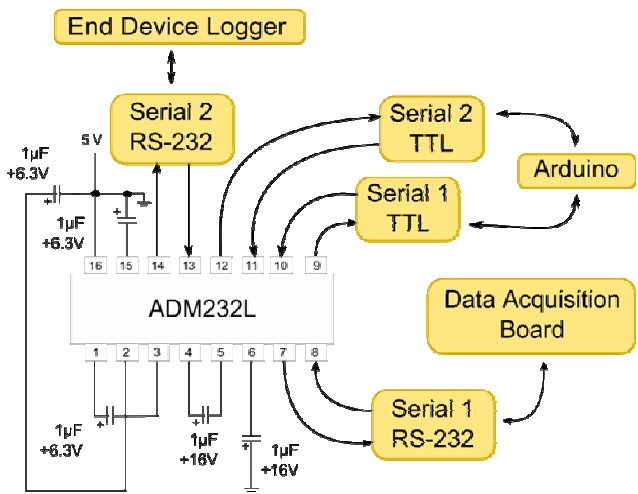

Fig. 11 TTL-to-RS232 Circuitry Scheme

4.5 End Device Cyclic Sleep

XBee End Devices support cyclic sleep, allowing the module to wake periodically to check for RF data and sleep when idle. Since the changes in the sampling frequency must be immediate (or almost immediate), checking incoming messages is performed every 28 seconds. However, XBee does not impose any restriction about this [20].

As the sample frequency rarely changes, XBee will not receive any data most of the times. In consequence, there is no need to wake the external circuitry whenever XBee awakes, so it makes sense to difference between XBee and external circuitry frequencies, $f_{XBee}$ and $f_{external}$, respectively. Equation 1 shows the relation between both variables:

$$f_{external} = N * f_{XBee} \qquad (1)$$

For example, consider an XBee module waking once every 28 seconds, and wakening an external sampling circuitry once every 2 minutes thought an interruption line. XBee must be set with the following parameters:

- $f_{external} = 120$
- $f_{XBee} = 28$
- $N = 120/28 \approx 4$

The Figure 12 represents graphically this example. The external circuitry is awake every four times XBee awakes and therefore matching the specified timing, as the timelines show. This behavior is repeated cyclically and hence its name.

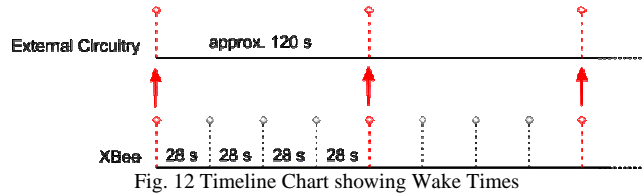

Fig. 12 Timeline Chart showing Wake Times

## 5. Testing

Several tests dealing with coverage and power consumption where done in order to evaluate the prototype performance.

5.1 Coverage Test

This test has consisted in detecting the mean RSSI (Received Signal Strength Indication) value, obtained

from the reception of 100 messages. Each measure is averaged from 5 repetitions of the same experiment, to counteract the signal fluctuations caused by indoor fading [32]. Both modules have transmitted with a power of 3 dBm and *boost* mode enabled [20]. Also, the ZigBee nodes have been configured to automatically select the channel having the least interference with other nearby networks [33]. The obtained results are shown in Tables 1 and 2.

| Free Space | Mean Attenuation (dB) |
|---|---|
| 50 cm | 0.00 |
| 1 m | 8.16 |
| 2 m | 11.65 |
| 4 m | 19.91 |
| 8 m | 23.93 |
| 11 m | 29.61 |

Table 1: Signal Loss in Free Space

| Obstacle | Mean Attenuation (dB) |
|---|---|
| Window (Open Metallic Blinds) | 1.04 |
| Window (Closed Metallic Blinds) | 3.95 |
| Wall with Open Door | 0.39 |
| Wall with Closed Door | 1.19 |
| Brick Wall | 1.46 |
| Between Floors | 13.08 |

Table 2: Signal Loss with Obstacles

Note that the total attenuation is the sum of free space and all obstacles loss [22].

## 5.2 Power Consumption Test

Prototype power consumption has been measured for each of the possible states using a polymeter:

| Node State | Consumption (mA) |
|---|---|
| Sleeping | 21.10 |
| Awake (Idle) | 69.80 |
| Awake (Transmitting) | 109.80 |

Table 3: Power Consumption Test

The consumption of a sleeping node is abnormally high, as shown in Table 3, due to a design fault in the Arduino board. This board always has the same consumption, regardless of whether the microcontroller and XBee are sleeping or not [34]. As a result, the authors are developing their own customized Arduino board where this problem is solved. On the other hand, it is important to highlight the fact that data sending causes a consumption peak, which, however high, lasts a very short time. The last measure, that corresponding to consumption when the node is awake and active, has been estimated considering transmit and every circuit component consumption.

## 5.3 Testing Scenario

In this example, a network of three nodes, one of each type, is deployed. As shown in Figure 13, the End Device is placed in Classroom 1.1 and the Router in the Repository, whilst Coordinator is in Laboratory 1.2.

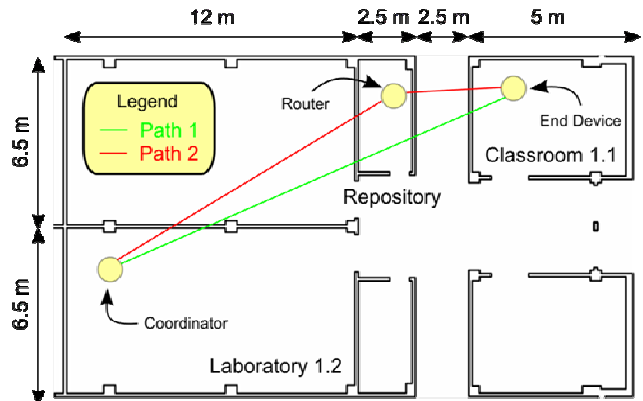

Fig. 13 Node distribution in the proposed Scenario

A temperature sensor is attached to the End Device (see Figure 14), sending temperature measures twice per hour. This rate was set from the Coordinator.

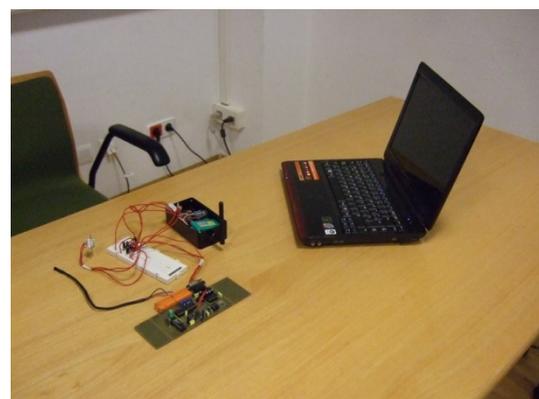

Fig. 14 End Device with a Temperature Sensor in Classroom 1.1

Unlike the previous nodes, the Coordinator is placed in the Laboratory 1.2, and it is connected to a computer through a USB port, as shown in Figure 15. The Coordinator Software is running on this computer, allowing the user to set the sample collecting frequency, or even store the samples in a file.

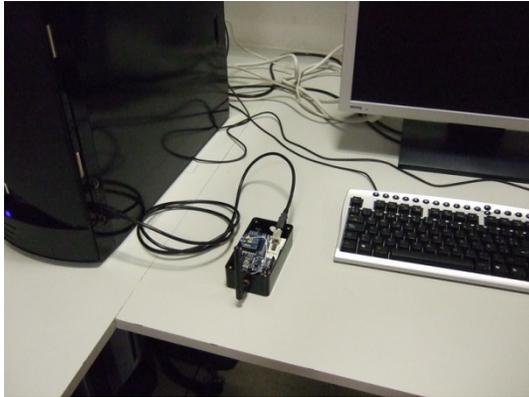

Fig. 15 Coordinator connected to a Computer USB Port in Lab 1.2

This scenario has been tested in real time, using a simple network with three nodes: One Coordinator, one Router and one End Device. When the Router is off, the received signal power (see Path 1 in Figure 13) at the Coordinator is too low to establish a connection with the End Device. However, when the Router located between the Coordinator and the End Device is switched on, the received signal power increases (it would be around 29.53 dBm, that is 3 dBm of transmit power, -29.61 dB of free space loss and -2.92 dB of 2-break wall obstacle loss), enabling communication between both of them.

With this scenario is possible to estimate End Device battery lifetime, as well. Considering a 30-minute sleeping cycle with 5-second staying in active state, the average consumption calculated according with the values shown in Table 3 is 23.79 mA. Therefore, using an 1100 mAh battery the total lifetime estimated is around 51 hours. Note this consumption is too high, caused by the Arduino board design, as it was said before. Therefore, the authors of this article are working in the design of a new Arduino-based board with very low consumption in sleeping state (just of a few μA). Thanks to this improved design, it is possible to extend battery lifetime to several months.

## 6. Conclusions

This article has presented a construction pathology detection system, based on a wireless sensor network using the ZigBee technology, which enables continuous monitoring of the parameters of interest, meeting the requirements of low consumption, ease of maintenance and installation flexibility. Its functional parts were fully discussed with diagrams, including the protocol specifically designed to collect samples from sensor nodes and several photos of the prototype built.

In addition, the results showing the typical node coverage limits and their consumption have been calculated for several situations.

**Acknowledgments**

This work has been supported by: 07TIC019105PR (Xunta de Galicia, Spain), TSI-020301-2008-2 (Ministerio de Industria, Turismo y Comercio, Spain) and 08TIC014CT (Instituto Tecnológico de Galicia, Spain).

**Santiago J. Barro Torres.** He obtained a MS in Computer Engineering in 2008 and a MS in Wireless Telecommunications in 2009, both from A Coruña University. His research interests are Digital Communications, Wireless Sensor Networks, Microcontroller Programming and RFID Systems.

**Carlos J. Escudero Cascón.** He obtained a MS in Telecommunications Engineering from Vigo University in 1991 and a PhD degree in Computer Engineering from Coruña University, in 1998. He obtained two grants to stay at Ohio State University as a research visitor, in 1996 and 1998. In 2000 he was appointed Associate Professor, and more recently, in 2009, Government Vice-Dean in the Computer Engineering Faculty at Coruña University. His research interests are Signal Processing, Digital Communications, Wireless Sensor Networks and Location Systems. He has published several technical papers in journals and congresses, and supervised one PhD Thesis.